**Maximum CME speed as an indicator of solar and geomagnetic activities**

A. Kilcik[1], V.B. Yurchyshyn[1], V. Abramenko[1], P.R. Goode[1],
N. Gopalswamy[2], A. Ozguc[3], J.P. Rozelot[4]

[1]*Big Bear Solar Observatory, Big Bear City, CA 92314, USA*
[2]*NASA Goddard Space Flight Center, Greenbelt, MD 20771, USA*
[3]*Kandilli Observatory and E.R.I Bogazici University, 34684, Turkey*
[4]*Nice University, OCA-Fizeau Dpt. Av. Copernic, 06130 Grasse, France*

**Abstract**

We investigate the relationship between the monthly averaged maximal speeds of coronal mass ejections (CMEs), international sunspot number (ISSN) and the geomagnetic Dst and Ap indices covering the 1996-2008 time interval (solar cycle 23). Our new findings are as follows. i) There is a noteworthy relationship between monthly averaged maximum CME speeds and sunspot numbers, Ap and Dst indices. Various peculiarities in the monthly Dst index are better correlated with the fine structures in the CME speed profile than that in the ISSN data. ii) Unlike the sunspot numbers, the CME speed index does not exhibit a double peak maximum. Instead, the CME speed profile peaks during the declining phase of solar cycle 23. Similar to the Ap index, both CME speed and the Dst indices lag behind the sunspot numbers by several months. iii) CME number shows a double peak similar to that seen in the sunspot numbers. The CME occurrence rate remained very high even near the minimum of the solar cycle 23, when both sunspot number and the CME average maximum speed were reaching their minimum values. iv) A well defined peak of the Ap index between May 2002 and August 2004 was co-temporal with the excess of the mid-latitude coronal holes during solar cycle 23. The above findings suggest that the CME speed index may be a useful indicator of both solar and geomagnetic activity. It may have advantages over the sunspot numbers, because it better reflects the intensity of Earth directed solar eruptions.

## 1. Introduction

When coronal mass ejections (CMEs) erupt from the Sun, high speed particles and strong magnetic fields can hurl earthward thus causing a significant impact on the near Earth space environment (geomagnetic storms) such as adverse effects on satellites and communications, electric power, pipelines, etc. Numerous severe storms occur during the maximum phase of the solar cycle, and they are mostly associated with CMEs (Gopalswamy et al. 2007, Zhang et al. 2007). Disturbances of the near Earth environment are measured by various parameters, such as *aa* (Mayaud 1972), Ap (Bartels et al. 1939) and Dst (Sugiuara 1964) indices, to name a few. Variations in solar activity are traced by measuring sunspot numbers (Hoyt & Schatten 1998), solar flare indices (Kilcik et al. 2010), and total solar irradiance (Lean et al. 1995). Gopalswamy (2006) introduced CME daily rate as a new solar activity indicator closely correlated to the geomagnetic activity. All these indices display correlative relationships with one another. Although the relationship between the solar and geomagnetic activity indices has been extensively studied (e.g., Stamper et al. 1999), it still eludes satisfactory explanation (Echer et al. 2004).

In this study, we use the linear CME speeds to further explore their geomagnetic activity. One obvious reason to use this parameter is that fast CMEs are very often associated with strong geomagnetic storms (Srivastava & Venkatakrishnan 2004, Yurchyshyn et al. 2004, 2005) and the correlation is the best when an earthward CME is associated with a magnetic cloud (Gopalswamy 2010). While sunspot numbers are quite suitable for characterizing solar activity, they may not always accurately reflect the overall intensity of solar eruptions, since not all sunspot groups are equally capable of producing powerful energetic events (Shi & Wang 1994, Abramenko 2005). There is a tendency to have more flares in the declining phase of a sunspot cycle and this tendency was even stronger during the decline of the $23^{rd}$ cycle (Bai 2006). The monthly maximum CME speed therefore is probably modulated by super active regions, which do not usually follow solar activity cycle (Gopalswamy et al. 2006). Moon et al. (2002), however, reported only a weak correlation between time integrated X-ray flux of CME associated flare and CME speeds. Therefore, the CME speed index as a measure of geo-effective solar activity may have advantages over the sunspot numbers in that it is more objective and better reflects the intensity of Earth directed solar eruptions.

Variations correlated with solar activity cycle were reported earlier for CME occurrence rate and speeds (Hildner et al. 1976, Webb & Howard 1994, Gopalswamy et al. 2003), latitude distribution (Gopalswamy et al. 2003), angular widths (Kahler et al. 1989, St. Cyr & Webb 1991). On the other hand, a number of papers analyzed the physical relationship between various CME parameters (speed, angular extent, orientation, rate, etc.) and geomagnetic storms (Richardson et al. 2002, Yurchyshyn et al. 2004), Kp index (Zhang et al. 2003, Miyoshi & Kataoka 2005), Ap index (Leamon et al. 2003, Forbes et al. 2005) and *aa* index (Luhmann 1997, Richardson et al. 2002). Many of these papers were either case studies or studies on small statistics not aimed at exploring a full solar cycle variation and long-term relationships. Gopalswamy et al. (2003) reported that the CME occurrence rate peaks two years after the solar cycle maximum. Ramesh (2010) further found that this lag is minimized when the sunspot area is used to describe the solar activity. Also, CMEs with higher speeds appear to follow the sunspot cycle much better then the entire population of CMEs.

Here we propose and explore a new solar activity index based on the maximum speed of CMEs (Section 2). In Section 3, we compare the CME speed index with the sunspot numbers and the geomagnetic Ap and Dst indices. The discussion and conclusions are given in Section 4.

## 2. Data and Methods

To derive the CME speed index (also monthly averaged maximum CME speeds), we used data from the Solar and Heliospheric Observatory (SOHO) mission's Large Angle and Spectrometric Coronagraph (LASCO, Brueckner et al., 1995) as compiled in the CME catalog[1] (Yashiro et al. 2004, Gopalswamy et al. 2009). The catalog covers the period from 1996 until the present. For each day, we only chose the eruption with the maximum linear fit speed, and the CME speed index was calculated as a monthly averaged maximum CME speed (Figure 1, solid line). A total of 3925 daily maximum speed measurements were selected out of a 4740 day interval and used for the correlation analysis. There were two large gaps in the CME data covering July-September 1998 and January 1999. CME rate was calculated from the CME number by counting all events reported for a given Carrington rotation (CR) and dividing the sum by the LASCO operational time over that CR (Gopalswamy et al. 2003). The CME rate is somewhat subjective, due to the difficulty in identifying the narrow and weak CMEs during solar maximum (Yashiro et al. 2008). For this reason Gopalswamy et al. (2010) considered CMEs with width exceeding 30 degree, which showed a good correspondence with sunspot numbers (see Fig. 2, 3 in Gopalswamy et al. 2010). In this paper the CME number is calculated based on all events in the CME catalog.

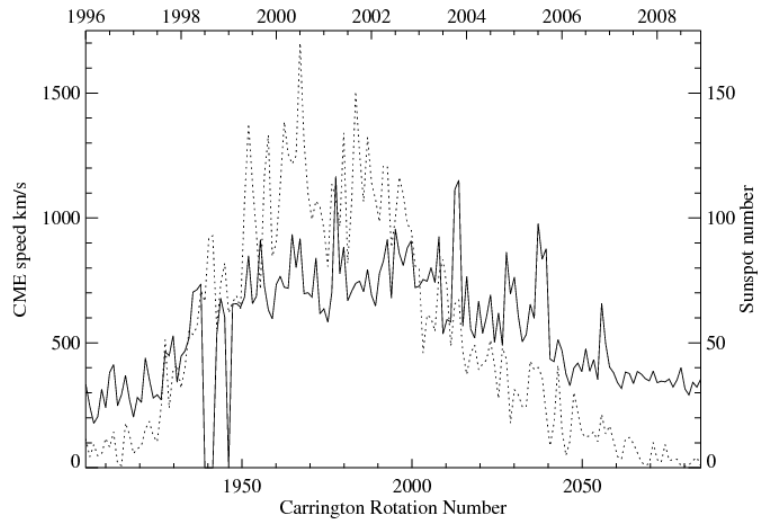

*Figure 1. Comparison of monthly data sets for solar cycle 23. For display purposes, ISSN (dotted line) values were scaled.*

The ISSN and the geomagnetic Ap index data were taken from the National Geophysical Data Center[2] for the investigated solar cycle 23. The Dst index data were provided by the World Data Center for Geomagnetism at Kyoto University[3] for the same time interval.

The planetary Ap index measures the solar particle effect on Earth's magnetic field, and characterizes the general level of geomagnetic activity over the globe for a given day. It is derived from *a* and Kp indices (Bartels et al. 1939), measured at a number of mid-latitude stations world-wide, characterizing variations of the geomagnetic field due to currents flowing in Earth's ionosphere and, to a lesser extent, in Earth's magnetosphere.

The hourly Dst index (Sugiura 1964) is obtained from several magnetometer stations near the equator. Dst index is a direct measure of the hourly averaged perturbation of the horizontal (H) component of the geomagnetic field caused by the varying magnetospheric ring current. Large negative Dst values indicate an increase in the intensity of the ring current (geomagnetic storm). Fares Saba et al. (2005) showed that the Ap and Dst indices are highly correlated during the geomagnetic storms mainly because in both cases the ring current is a dominant contributor. Figure 2 plots monthly averaged values of the Ap and Dst indices. Although their absolute values differ, at times significantly, there is general agreement (correlation coefficient is -0.81) and synchronous peaks and troughs are apparent.

---

[1] http://cdaw.gsfc.nasa.gov/CME_list/index.html
[2] http://www.ngdc.noaa.gov/
[3] http://wdc.kugi.kyoto-u.ac.jp/dstdir/

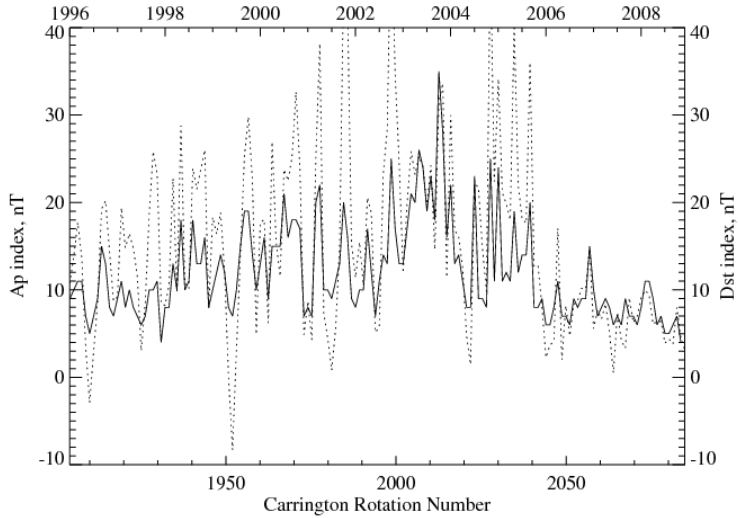

*Figure 2. Comparison of monthly geomagnetic Ap and Dst (dotted) indices. For display purposes the sign of the Dst index was reversed.*

We began analysis with checking the existence of autocorrelation, which detects randomness in a data set by computing autocorrelations for varying time lags. If a data set is random, autocorrelation is nearly zero for any and all time-lag separations. In case of a non random time series, one or more of autocorrelations will be significantly non-zero (Box & Jenkins 1976). To explore the autocorrelation we applied Durbin-Watson test statistic used to detect the presence of autocorrelation. The test based on the residuals, $x_t$, between observed and predicted data and test statistic is given by

$$d = \frac{\sum_{t=2}^{T}(x_t - x_{t-1})^2}{\sum_{t=1}^{T} x_t^2} \qquad (1)$$

where $t$ is time of observation. The test statistic, $d$, ranges from 0 to 4. Values of d=2 indicate the absence of autocorrelation, while values smaller than 2 suggest positive autocorrelation, and test results exceeding 2 are evidence of negative autocorrelation (Durbin & Watson 1951).

To investigate the relationship between maximum CME speed and other data sets used here we used the cross-correlation analysis, which determines correlation coefficient, $r$, between two time series with a possible time delays.

## 3. Results

Figure 1 plots CME speed index and ISSN for the investigated solar cycle 23. The CME speed index shows the presence of the 11 year solar activity cycle. In general, the monthly means of daily maximum speeds exceed 700 km/s during the maximum phase, while they remain on average at a ~350 km/s level during the rising and declining phase of the solar cycle. We emphasize that the monthly averaged maximum CME speeds do not exactly follow the ISSN during the solar cycle 23. During the rising phase of the cycle both indices show similar behavior, however, the maximum and declining phases reveal differences between them. The double maximum in the ISSN is not at all prominent in the speed index. Instead, the CME speed index gradually rises until it peaks at CR 1995 (October 2002). While the sunspot number rapidly drops in the declining phase, the monthly averaged speed remains relatively high, even when there are only few sunspots. As it follows from the figure, the maximum of the CME speed index appears to be delayed relative to the 23$^{rd}$ solar cycle maximum (October 2000) by nearly two years.

The spikes seen in Figure 1 are also present in the plot of the mean speed reported in Gopalswamy et al. (2010). It must be pointed out that the mean CME speed (see also Gopalswamy et al. 2003, 2009) differs from the CME index in the present work. The mean speed includes the speeds of *all* CMEs averaged over Carrington Rotation periods. The mean speed follows the ISSN and shows a double hump with a dip during the solar maximum when plotted as annual averages (year 2001, see Fig. 1 in Gopalswamy et al. 2008b). To the contrary, the CME maximal speed index, studied here, is based on the *highest daily* CME speed measurement averaged over a month. We thus select only the maximum speeds, so the super active regions get higher weight because they produce high speed CMEs in greater numbers (Gopalswamy et al. 2007, Kilcik et al. 2010). As a result of this selection, the CME maximal speed index matches the number of fast and wide CMEs reported in Gopalswamy et al. (2008b). Since the geo-effective population of CMEs is generally fast and wide, we conclude the CME speed index may be a good indicator of the geo-effectiveness as will be shown below.

To test the CME speed index randomness we used the Durbin-Watson test described above. This test makes use of the residual difference between observed and predicted values of a given parameter. The predicted CME speed index was generated from the regression equation (2) that relates the observed monthly maximum CME speeds and the ISSN for solar cycle 23:

$$V_{CME\,max}^{p} = 3.769 \times ISSN + 376.08 . \tag{2}$$

Similarly we obtained predicted values for the geomagnetic Ap and Dst indices using the following equations:

$$Ap^{p} = 0.0174 \times V_{CME\,max}^{O} + 1.9178 \tag{3}$$

$$Dst^{p} = -0.0258 \times V_{CME\,max}^{O} - 0.8888 . \tag{4}$$

In these equations the superscripts *P* and *O* denote the predicted and observed values respectively.

The Durbin–Watson test applied to the data showed that the maximum CME speed, geomagnetic Ap, and Dst indices display positive autocorrelation with the test statistic $d_{speed}$=1.018, $d_{Ap}$=1.2, and $d_{Dst}$=1.19 respectively. We thus conclude that these indices are not random data sets.

To quantify the relationship among various indices, we made use of the cross-correlation analysis and found good correlations between the observed CME speed index and ISSN (r = 0.76±0.09), Ap index (r = 0.68±0.1) and the Dst index (r = -0.53±0.13) (also Figure 3 upper row, and Table 1). We would like to point out that the highest correlation coefficients are found for zero time lags. To estimate the error level of the cross-correlation coefficients, we applied the Fisher's test[4], gave us upper and lower bounds of the confidence level for the correlation coefficient (horizontal dashed lines in Figure 3, and Table 1).

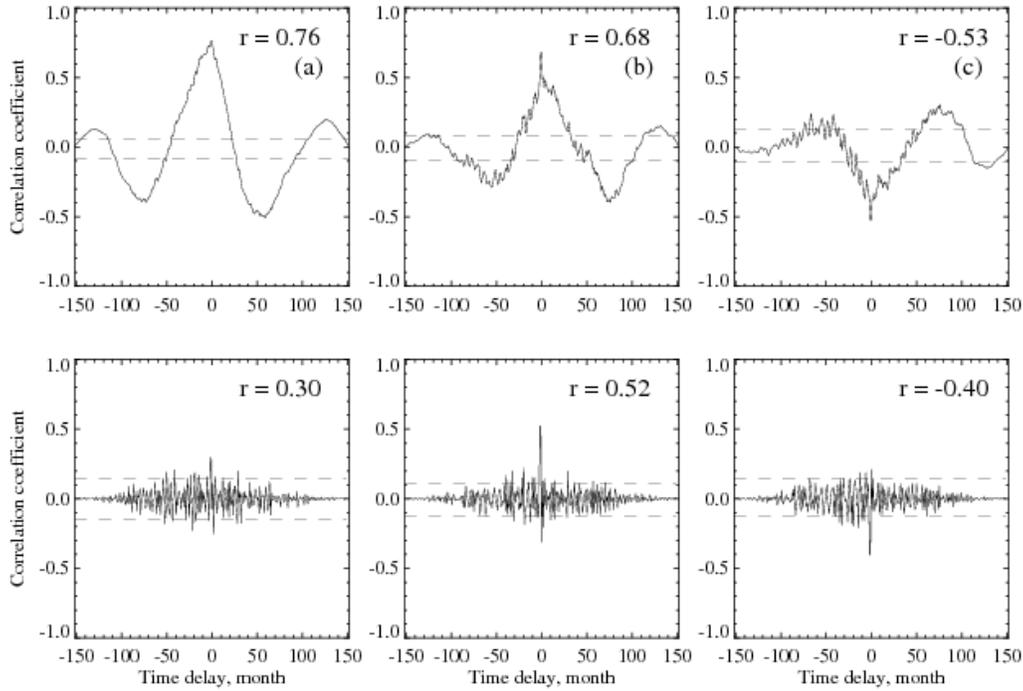

*Figure 3. Results of cross-correlation analysis between the CME speed index and the ISSN (left panels), Ap index (middle panels), and the Dst index (right panels). The upper raw of panels show results for observed data, while the bottom panels show results for detrended data sets. The error levels were calculated by using the Fisher's test.*

To remove the contribution of the solar cycle (general trend) into the above cross-correlation analysis and to determine the main contributor to the correlations, we also tested cross-correlation for detrended data. Detrending was achieved by generating differenced data sets, i.e., by calculating difference between two neighboring monthly values in a time series. The detrended analysis show that the high correlation between the maximum CME speed and the ISSN is mainly due to the 11-year solar cycle (the detrended correlation strongly decreased). Contrary to this, the detrended cross-correlation coefficients between the maximum CME speed and both geomagnetic indices decreased only slightly (by about 25 %), indicating that the main contributor to their correlation are fluctuations in 11-year solar activity cycle (see Figure 3, lower panels).

---

[4] http://icp.giss.nasa.gov/education/statistics/

*Table 1.* *Results of the cross-correlation analysis. The values in brackets show detrended correction coefficients.*

|     | ISSN              | CME speed index       |
| --- | ----------------- | --------------------- |
| Ap  | 0.51±0.13 ( 0.15) | 0.68±0.10 ( 0.52±0.13) |
| Dst | -0.37    (-0.08)  | -0.53±0.13 (-0.40)    |

As it follows from Table 1, CME speed index shows higher correlation with Ap and Dst indices than ISSN does. Therefore, we argue that the CME speed index is sensitive to both geomagnetic activity and solar activity. Also, the CME speed index and the Dst index show the lowest correlation (see Table 1 and Figure 3), which could be due to a very irregular time profile of the Dst index (as compared to the Ap index). The lack of correlations may come from two sources: 1) some strong eruptions were not directed toward the Earth thus did not cause any response in the Dst index, while contributing to the CME speed index and 2) some CMEs may had unfavorable orientation of magnetic fields and did not cause strong geomagnetic storms. This can be also seen from a very strong correlation between the Dst index and the speeds of CMEs, associated with magnetic cloud structures (Burlaga et al. 1981; Yurchyshyn et al. 2004, 2005; Gopalswamy et al. 2008a). However, this correlation is substantially weaker when only non-magnetic cloud ejecta are considered (Gopalswamy 2010).

Next we will explore whether this index may help us better understand the details of solar-terrestrial interactions from one cyclic data analysis. Figure 4 plots 12 step running averages of the ISSN, CME speed and numbers as well as the Ap and Dst indices for the solar cycle 23. There are several moments that we would like to point out.

First, the plot shows that the peak of the averaged maximum CME speed (CR 1995, October 2002) and the local peak in the CME number (CR 1993, August 2002) are delayed relative to the second sunspot maximum (CR 1985, January 2002). Earlier, Gopalswamy et al. (2009); and Ramesh (2010) reported that the CME occurrence rate similarly lags behind the sunspot maximum.

Second, there is a general trend for the CME speeds, CME numbers and sunspot numbers to behave quite differently in the declining phase of the 23$^{rd}$ solar cycle. After October 2002 (i.e., CR 1995), the average maximum speed began to decline and the trend of the decline is similar to that displayed by the Dst index and is also consistent with the sunspot numbers. However, in year 2004 both the CME speed and Dst indices began to increase again and their values peaked during in the middle of year 2005 (CR 2035), all while the sunspot number continued to gradually vanish.

Another interesting feature in this plot is an increase in the CME number during the solar minimum (CR 2055-2075). This may be an artifact of the subjective CME counting: faint narrow CMEs were easily detected during the declining phase than during the maximum phase (see Yashiro et al. 2008). This disparate behavior can also be explained by the fact that a significant fraction of CMEs at this time could be low speed events related to quiescent and high latitude polar crown filaments, which are not associated with sunspots (Gopalswamy et al. 2010, Ramesh 2010).

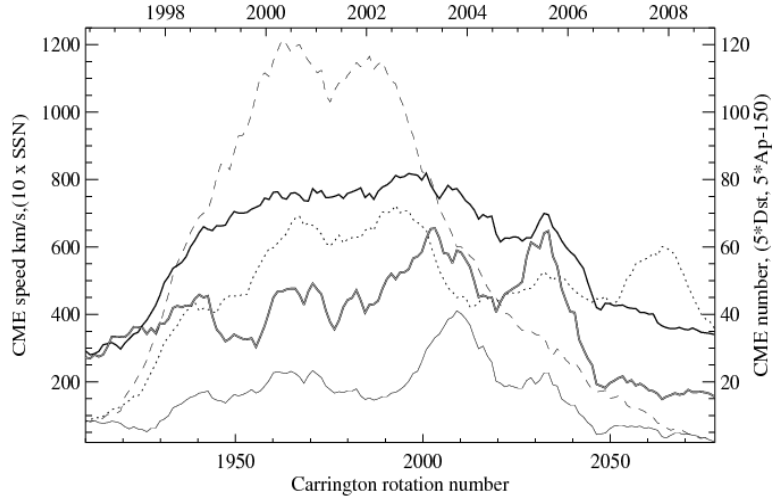

*Figure 4. Time profiles of 12 point running averaged monthly data sets. In this plot the dashed line shows the sunspot numbers, the bold solid line - the CME index, the dotted line is the CME number, the double line represents the Dst index, and the thin solid line is the Ap index.*

The third point is related to the cause of the burst of geomagnetic activity in 2003. The plots clearly show that according to one solar cycle data analysis results the Dst and Ap indices are mainly defined by different drivers, although some CME activity is clearly imprinted, at times, onto the Ap profile (e.g., compare the CME number and Ap profiles between CR1955 and CR1975). The Dst index peaks nearly simultaneously with the CME speeds, while the Ap index reached its maximum at the time (CR 2010) when the number of CME drops, while coronal holes (CHs) observed on the solar disk peaked (Abramenko et al. 2010). We thus argue that both CME and CH activity leaked into each of the geomagnetic indices, although in different degrees, which may explain the lag between the Dst and Ap indices.

## 4. Discussion and Conclusions

Our analysis showed that all analyzed data sets have significant relationships with each other. This study based on the comparison of CME speed index, sunspot numbers, geomagnetic Ap and Dst indices confirms that the CME speed index may be a useful index that simultaneously measures solar and geomagnetic activity. Further studies are needed to explore the details of the relationship between this new index and geomagnetic effects.

We summarize the main findings of this study as follows:
1) There is a good relationship between monthly averaged maximum CME speeds and sunspot numbers, Ap and Dst indices (correlation coefficients are 0.76, 0.68, -0.53 respectively). We note that the CME speed index displays better correlation with the geomagnetic indices than the sunspot numbers, which suggests that the CME speed index may be a powerful indicator of both solar and geomagnetic activity.

2) Unlike the sunspot numbers, the CME speed index does not exhibit a double peak maximum. Instead, the CME speed profile peaks during the declining phase of solar cycle 23.
3) CME number shows a double peak similar to that seen in the sunspot numbers. The CME occurrence rate remained very high even near the minimum of the solar cycle 23, when both sunspot number and the CME average maximum speed were reaching their minimum values.
4) A well defined peak of the Ap index between May 2002 and August 2004 was co-temporal with the excess of the mid-latitude coronal holes during solar cycle 23.

We start the discussion by noting that Feynman (1982) decomposed the annual average *aa* index into two periodic functions and found that they are nearly 180 degrees out of phase. The first component, synchronized with the sunspot cycle, was proposed to be due to solar flares, CMEs and transient coronal holes. The other component lags behind the sunspot cycle and peaks during the declining phase. The authors proposed that this component of geomagnetic activity is due to long lived solar wind sources such as polar coronal holes. Echer et al. (2004) analyzed the lag between solar activity, as measured by sunspot numbers, and the *aa* geomagnetic index. The data set covered period between 1868 and 2000. These authors concluded that the lag varies from one cycle to another, reaching 2 year for cycle 22. The explanation for the lag is in the dual peak structure in the *aa* index. The first peak is related to sunspots' CME activity and the second peak is thought to be caused by fast solar wind streams, which increase during the declining phase of the solar cycle as more and more mid-latitude coronal holes appear on the solar surface (Legrand & Simon 1985). Indeed, according to Abramenko et al. (2010), the declining phase of the $23^{rd}$ solar cycle displayed an excess of low-latitude coronal holes. While our analysis supports the Echer et al. (2004) report, the CME speed index, newly introduced, in this study may refine the explanation for the cause of the second peak in the geomagnetic activity. We thus propose that since the peak of CME speeds is nearly co-temporal with the peak in the Ap index, fast CME may also be responsible for the burst of geomagnetic activity during a two-year period starting in May 2002. To further elaborate this idea Kilcik et al. (2010) analyzed time distribution of small and large sunspot groups and concluded that during the solar cycle 23 the number of large groups peaks about two years after the maximum time of the ISSN and small groups. Thus the enhanced CME maximal speed index at the declining phase of the solar cycle may reflect on the excess of large and complex active regions during this period of time.

Another aspect we would like to emphasize is that there is a significant correlation between the monthly averaged maximum CME speed and the geomagnetic indices, and this correlation mainly comes not from the cyclic dependence but from the fluctuations in CME speed and geomagnetic indices. This finding, based on monthly averaged data, is in line with earlier works (e.g., Srivastava & Venkatakrishnan 2004 and Yurchyshyn et al. 2004, 2005), which were based on the analysis of individual events. Thus, Srivastava & Venkatakrishnan (2004) selected 64 geo-effective CME events and found a good negative correlation (-0.66) between CME speeds and the corresponding Dst index. Yurchyshyn et al. (2004, 2005) concluded that the CME speed is directly related the intensity of the Bz component of the associated magnetic clouds (also see Qiu & Yurchyshyn 2005), which, in turn, affects the magnitude of the Dst index. Gopalswamy (2010) examined the relationship between CME speeds and the Dst index separately for different types of ejecta. The CMEs, exhibiting signatures of magnetic clouds, show the best correlative relationship between speed and the Dst index, because these CMEs erupt from the disc and hence head directly to the Earth.

And finally, sunspot numbers are not the best indicator of solar activity, since they do not contain information on how magnetically energetic sunspot regions are. During the solar minimum, when the sunspots vanish, the sunspot numbers cease to provide any information about what to expect at the Earth at all. In contrast, the CME maximum speed index provides information about energy of solar events, not just the frequency of solar magnetic activity. We, thus, exploit the physical link between CMEs and geomagnetism and present a new and more meaningful index to describe solar geoeffectiveness.

We would like to thank referees for valuable comments and suggestions, which led to a significant improvement of the paper. We acknowledge usage of ISSN and Ap index from National Geophysical Data Center. The Dst index data was provided by the World Data Center for Geomagnetism at Kyoto University. The CME catalog is generated, and maintained by the Center for Solar Physics and Space Weather, the Catholic University of America in cooperation with the Naval Research Laboratory and NASA. SOHO is a project of international cooperation between ESA and NASA. We thank W. Cao for help during the manuscript preparation. This research was supported by NASA grants GI NNX08AJ20G and LWS NNX08AQ89G as well as NSF ATM0716512 grant.